\documentclass[preprint, %
aps,amsmath,amssymb,prx,
superscriptaddress, %
longbibliography, %
floatfix]
{revtex4-2} %

\usepackage{graphicx} %
\usepackage{siunitx}
\usepackage{hyperref}

\date{October 29, 2025}

\begin{document}

\title{Second-order Stark shifts exceeding 10$\,$GHz in electrically contacted SiV$^-$ centers in diamond}

\author{Manuel Rieger}
\affiliation{%
 Walter Schottky Institute and School of Natural Sciences,
 Technical University of Munich, 85748 Garching, Germany
}%
\affiliation{%
 Munich Center for Quantum Science and Technology (MCQST), 80799 Munich, Germany
}%

\author{Nori N. Chavira Leal}
\affiliation{%
 Munich Center for Quantum Science and Technology (MCQST), 80799 Munich, Germany
}%
\affiliation{%
 Walter Schottky Institute and School of Computation, Information and Technology,
 Technical University of Munich, 85748 Garching, Germany
}%

\author{Rubek Poudel}
\affiliation{%
 Walter Schottky Institute and School of Natural Sciences,
 Technical University of Munich, 85748 Garching, Germany
}%
\affiliation{%
 Munich Center for Quantum Science and Technology (MCQST), 80799 Munich, Germany
}%

\author{Tobias Waldmann}
\affiliation{%
 Munich Center for Quantum Science and Technology (MCQST), 80799 Munich, Germany
}%
\affiliation{%
 Walter Schottky Institute and School of Computation, Information and Technology,
 Technical University of Munich, 85748 Garching, Germany
}%

\author{Lina M. Todenhagen}
\affiliation{%
 Walter Schottky Institute and School of Natural Sciences,
 Technical University of Munich, 85748 Garching, Germany
}%
\affiliation{%
 Munich Center for Quantum Science and Technology (MCQST), 80799 Munich, Germany
}%

\author{Stefan Kresta}
\affiliation{%
 Munich Center for Quantum Science and Technology (MCQST), 80799 Munich, Germany
}%
\affiliation{%
 Walter Schottky Institute and School of Computation, Information and Technology,
 Technical University of Munich, 85748 Garching, Germany
}%

\author{Viviana Villafañe}
\affiliation{%
 Munich Center for Quantum Science and Technology (MCQST), 80799 Munich, Germany
}%
\affiliation{%
 Walter Schottky Institute and School of Computation, Information and Technology,
 Technical University of Munich, 85748 Garching, Germany
}%

\author{Martin S. Brandt}
\affiliation{%
 Walter Schottky Institute and School of Natural Sciences,
 Technical University of Munich, 85748 Garching, Germany
}%
\affiliation{%
 Munich Center for Quantum Science and Technology (MCQST), 80799 Munich, Germany
}%

\author{Kai Müller}
\affiliation{%
 Munich Center for Quantum Science and Technology (MCQST), 80799 Munich, Germany
}%
\affiliation{%
 Walter Schottky Institute and School of Computation, Information and Technology,
 Technical University of Munich, 85748 Garching, Germany
}%

\author{Jonathan J. Finley}
\affiliation{%
 Walter Schottky Institute and School of Natural Sciences,
 Technical University of Munich, 85748 Garching, Germany
}%
\affiliation{%
 Munich Center for Quantum Science and Technology (MCQST), 80799 Munich, Germany
}%

\begin{abstract}
Negatively charged silicon vacancy centers (SiV$^-$) in diamond exhibit excellent spin coherence and optical properties, making them promising candidates for quantum technologies. However, the strain-induced inhomogeneous distribution of optical transition frequencies poses a challenge for scalability. We demonstrate electrical tuning of the SiV$^-$ center zero-phonon lines using in-plane contacts to apply moderate electric fields up to \SI{45}{\mega\volt/\meter}. The second-order Stark shift exceeds \SI{10}{\giga\hertz}, which is of the same order of magnitude as the \SI{15}{\giga\hertz} inhomogeneous distribution of SiV$^-$ observed in emitters embedded in optical nanostructures such as photonic crystal nanocavities. Analysis of individual SiV$^-$ centers shows significant variation in polarizabilities between defects indicating that the polarizability strongly depends on local parameters like strain. The observed polarizabilities are $\sim$3--25 times larger than those of tin vacancy centers, which we attribute to valence band resonances that delocalize the $e_u$ wavefunctions. Photoluminescence excitation measurements reveal that optical linewidths increase moderately with applied electric field strength. Our results demonstrate that large electrical Stark shifts can overcome the inhomogeneous distribution of transition frequencies, representing a significant step toward scalable SiV$^-$-based quantum technologies such as quantum repeaters.
\end{abstract}

\maketitle

\section{Introduction}

Many quantum technologies require long coherence times for their quantum memories and efficient optical interfaces to send information over long distances \cite{Briegel1998_quantumRepeaters,Kimble2008_quantumInternetReview}. In this context, color centers in diamond are promising spin memories and spin-photon interfaces and have already been integrated into electrical and optical nanostructures \cite{Wan2020_Englund_chiplets,Knaut2024_Lukin_40kmEntanglement}. 
In particular, the inversion-symmetric group-IV vacancies in diamond, including silicon, germanium and tin vacancies, have excellent optical properties such as near-transform-limited zero-phonon lines \cite{Goerlitz_2020_transformLimitedLines}, electron spin coherence times up to milliseconds \cite{sukachev2017silicon_10msCoherence,Karapatzakis2024_tinVacancy_msCoherence} and nuclear spin coherence times exceeding seconds \cite{grimm2025coherentcontrollonglivednuclear_Jelezko2025_secondCoherence}.
Due to local strain and Coulomb fields caused by other defects in the crystal, the emission wavelength can vary slightly between emitters, making emitted photons distinguishable. Specifically, the inhomogeneous distribution of optical emission frequencies lies in the range of \SI{15}{\giga\hertz} for implanted SiV$^-$ incorporated into nanostructures, even after careful annealing \cite{Evans2016_annealingProcedure}. Overcoming this inhomogeneous distribution of optical emission frequencies is a practical challenge when building systems with multiple optically interfaced color centers.

Currently, different strategies are under investigation for making emitted photons indistinguishable: 1) When converting distinguishable SiV$^-$ photons to telecom frequencies, the target wavelength can be precisely tuned by adjusting the frequency of the pump laser used in the nonlinear frequency conversion process individually for each emitter \cite{Schfer2025_telecomConversion_SiV,Knaut2024_Lukin_40kmEntanglement}. 2) Strain can change the optical transition frequency and can be applied locally using advanced nanostructures such as electrically tunable cantilevers \cite{Meesala2018_SiV_strainTuning,Sohn2018_SiV_strainTuning_coherence,Li2024_strainTuning_tinVac,Maity2018_GeV_strainTuning_Loncar}. 3) Electric fields can tune the optical emission frequency via the DC Stark effect, an approach that has already been demonstrated for nitrogen and tin vacancy centers \cite{Sipahigil2012_NV_TPI_Lukin_StarkTuning,Bernien2012_Hanson_NV_StarkTuning,Aghaeimeibodi2021_Vuckovic_TinVac2021,DeSantis2021_Englund_StarkTun_TinVac_2021,bushmakin2025twophotoninterferencephotonsremote_Wrachtrup_tinVacancy}. 

Here, we investigate Stark tuning of SiV$^-$ centers, which provides straightforward electrical tunability with simple electrode structures. Our results show that the polarizability is 3--25 times larger for silicon vacancies compared to tin vacancies \cite{Aghaeimeibodi2021_Vuckovic_TinVac2021,DeSantis2021_Englund_StarkTun_TinVac_2021,bushmakin2025twophotoninterferencephotonsremote_Wrachtrup_tinVacancy}. Due to our relatively large \SI{7.6}{\micro\meter} electrode spacing, the applied electric fields are far from the limit posed by electrical breakdown in vacuum. Our finding that we can tune optical emission frequencies by multiple \SI{}{\giga\hertz} even with these moderate electric fields relaxes the constraints on nanostructure fabrication for overcoming inhomogeneous optical emission frequencies.

\section{Results}

\begin{figure*}[!htb]
\centering
\includegraphics[scale=1]{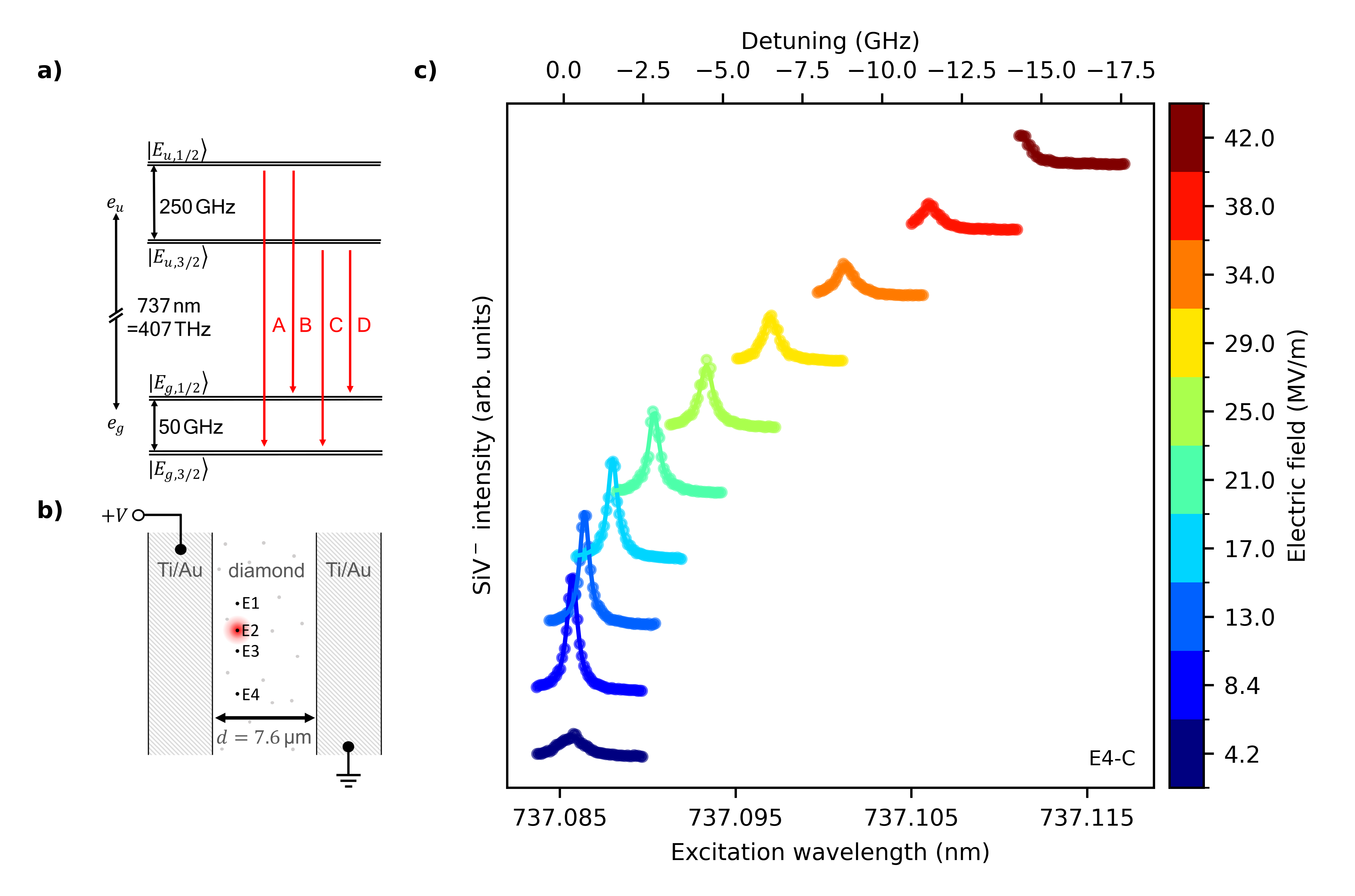}
\caption{\label{fig:Fig1}
\textbf{ SiV$^-$ energy level diagram, sample schematic and photoluminescence excitation spectra for a varying applied local electric fields. } 
a) Energy level diagram of SiV$^-$ centers. The ground states $|E_{g\text{1/2}} \rangle$ and $|E_{g\text{,3/2}}\rangle$, that have a hole in the $e_g$ states, are split by approximately \SI{50}{\giga\hertz} due to spin-orbit coupling and the Jahn-Teller effect. Strain can increase this splitting. The excited states $|E_{u\text{1/2}}\rangle$ and $|E_{u\text{3/2}}\rangle$, that both have a hole in the $e_u$ orbitals are lying close to the valence band maximum and split by at least \SI{250}{\giga\hertz}. The red arrows define the optical transitions A to D that can be observed in photoluminescence (PL) and PL excitation.
b) Top-view schematic of the relevant parts of the sample showing two thin Ti/Au electrodes on flat bulk diamond. The electrode distance is \SI{7.6}{\micro\meter} and we ground one of the electrodes and apply a voltage to the other. We focus the excitation and detection spot at approximately a quarter of the electrode distance from one of the electrodes. This breaks the symmetry in voltage because the photocurrent is stronger when the electrode closer to the excitation spot is positively biased. The small black spots E1, E2, E3 and E4 schematically mark different studied single SiV$^-$ positions along the electrode edges. The grey spots mark other single SiV$^-$, which we did not select.
c) Photoluminescence excitation spectra of the emitter labelled E4, transition C (denoted E4-C), for different applied electric fields show that the optical emission frequency of SiV$^-$ changes with the field. The solid lines represent Lorentzian peak fits of the data, individually performed for each voltage.
}
\end{figure*}

Figure \ref{fig:Fig1} a) depicts the simplified energy level diagram of SiV$^-$ in the absence of external magnetic fields \cite{GaliMaze,Thiering2018_MagnetoOpticalSpectra_G4V}. The ground states $|E_{g\text{,1/2}} \rangle$ and $|E_{g\text{,3/2}}\rangle$ are split by approximately \SI{50}{\giga\hertz} due to spin-orbit coupling and the Jahn-Teller effect, a splitting that can increase with strain. These states are defined by the occupation of the energy levels $e_u$ and $e_g$. The states $e_u$ are close to the valence band maximum, while the energetically higher-lying states $e_g$ are within the bandgap. In the ground states $|E_{g\text{,1/2}} \rangle$ and $|E_{g\text{,3/2}}\rangle$ of the SiV$^-$, the orbitals $e_u$ are filled entirely with 4 electrons, while there are 3 electrons or effectively a single hole in $e_g$. For the excited states $|E_{u\text{,1/2}}\rangle$ and $|E_{u\text{,3/2}}\rangle$, the situation is reversed, meaning that there is a hole in the $e_u$ levels \cite{GaliMaze}. The excited states are split by at least \SI{250}{\giga\hertz} and their splitting can also increase with strain. 
The energy of the orbital states shift with electric field as a consequence of the DC Stark effect \cite{DCStarkEffect_Henry_1965}. Due to the inversion symmetry of the atomic arrangement of a silicon vacancy complex \cite{Rogers2014,GaliMaze}, all SiV$^-$ states are associated with inversion-symmetric wavefunctions. This leads to a vanishing static electric dipole moment and thus the absence of a first order Stark shift. However, the negatively charged electron wavefunctions and positively charged nuclei can be polarized relative to each other by an applied electric field, leading to an induced electric dipole moment and a second-order shift of the electronic states energies. Hence, no linear Stark shift is expected, but second order Stark shifts can occur.

The Stark shift in the optical transition frequencies represents the difference of the Stark shifts of the participating initial and final states.
We describe the optical transition frequency
\begin{align}
    f(E) = f_{\text{max}} - \alpha (E-E_0)^2  \quad
    \label{eq:tuning1}
\end{align}
and Stark shift
\begin{align}
     \Delta f = f(E) - f_{\text{max}}.
    \label{eq:tuning2}
\end{align}
as a function of a local applied electric field $E$. Here,
$f_{\text{max}}$ is the maximum optical transition frequency, $E$ is the applied local electric field, $E_0$ is the local electric field offset and $\alpha$ is the effective polarizability of the transition frequency. The effective polarizability $\alpha$ is the difference of polarizabilities of the two levels involved in the optical transition, where each of those is a measure of the field-induced perturbation of the wavefunction. 

In this study, we use interdigitated electrodes on a flat oxygen-terminated diamond surface to apply an electric field to the silicon vacancies studied. Figure \ref{fig:Fig1} b) shows a schematic illustration of the electrodes and the measurement setting. The parallel electrode edges are separated by \SI{7.6}{\micro\meter}. One of the electrodes is grounded, while the positive and negative contacts of the voltage source are floating to avoid ground loops. We position the focal spots from the excitation and detection channels at a quarter of the electrode distance. Since we keep this position distance constant, we have a constant conversion factor between applied voltage and applied electric field, see details below.
For all measurements, we cool the sample to \SI{5}{\kelvin} in vacuum using a closed-cycle cryostat.

To estimate the local electric field experienced by the SiV centers for a given applied voltage, we simulate the system using the finite-element tool COMSOL and the Lorentz local field approximation. Details on the COMSOL simulation can be found in the Supplementary Information. The Lorentz approximation accounts for the polarization of a medium surrounding a color center, which effectively enhances the electric field at the defect site \cite{Tamarat2006}. The local electric field acting on the defect is then given by $E = E_{\text{ext}}(\varepsilon + 2)/3$, where $E_{\text{ext}}$ is the externally applied field obtained from the COMSOL simulations and $\varepsilon$ is the dielectric constant of diamond ($\varepsilon = 5.7$) \cite{DeSantis2021_Englund_StarkTun_TinVac_2021}. For an emitter \SI{100}{\nano\meter} below the diamond surface and located \SI{1.9}{\micro\meter} (quarter of the electrode separation) from an electrode with an applied voltage of \SI{10}{\volt}, the simulation gives $E_{\text{ext}} = \SI{0.82}{\mega\volt\per\meter}$. This yields a local field $E = 0.82 \times (5.7 + 2)/3 = \SI{2.10}{\mega\volt\per\meter}$. The uncertainty in the external field, based on assumed variations of the emitter position of \SI{1.5}{\micro\meter} to \SI{2.4}{\micro\meter} from the electrode, translates to a 7\% relative uncertainty in the local field estimation.

To locate SiV$^-$ centers, we translate the laser spot parallel to the electrode edge and measure the photoluminescence (PL) using a blue or green laser with either \SI{740}{\tera\hertz}, \SI{582}{\tera\hertz} or \SI{564}{\tera\hertz} optical frequency and an optical power of \SIrange{0.5}{3}{\milli\watt} for excitation. 
The $1/e^2$ beam waist is approximately \SI{600}{\nano\meter}.
A representative PL spectrum is displayed in the supporting information in Figure \ref{fig:FigPL}. For the given SiV$^-$ density, we always find several SiV$^-$ centers per detection volume, but they can be readily isolated spectrally by resonantly driving individual transitions. 

To strongly increase the spectral resolution and thereby effectively work with single SiV$^-$ centers, we measure PL excitation spectra in the following. For this, we scan the optical frequency of a narrow-linewidth ($<\SI{2}{\mega\hertz}$) wavelength-stabilized laser over the peaks of the photoluminescence spectra. A \SI{375}{} to \SI{400}{\tera\hertz} bandpass filter isolates the phonon sideband emission while suppressing the excitation lasers. To detect the emitted single photons we use single photon avalanche diodes or superconducting nanowire single photon detectors. To stabilize the SiV in the negative state SiV$^-$, we use an off-resonant low power green or blue laser (again \SI{740}{\tera\hertz}, \SI{582}{\tera\hertz} or \SI{564}{\tera\hertz} optical frequency).

\begin{figure}[!htb]
\centering
\includegraphics[scale=1]{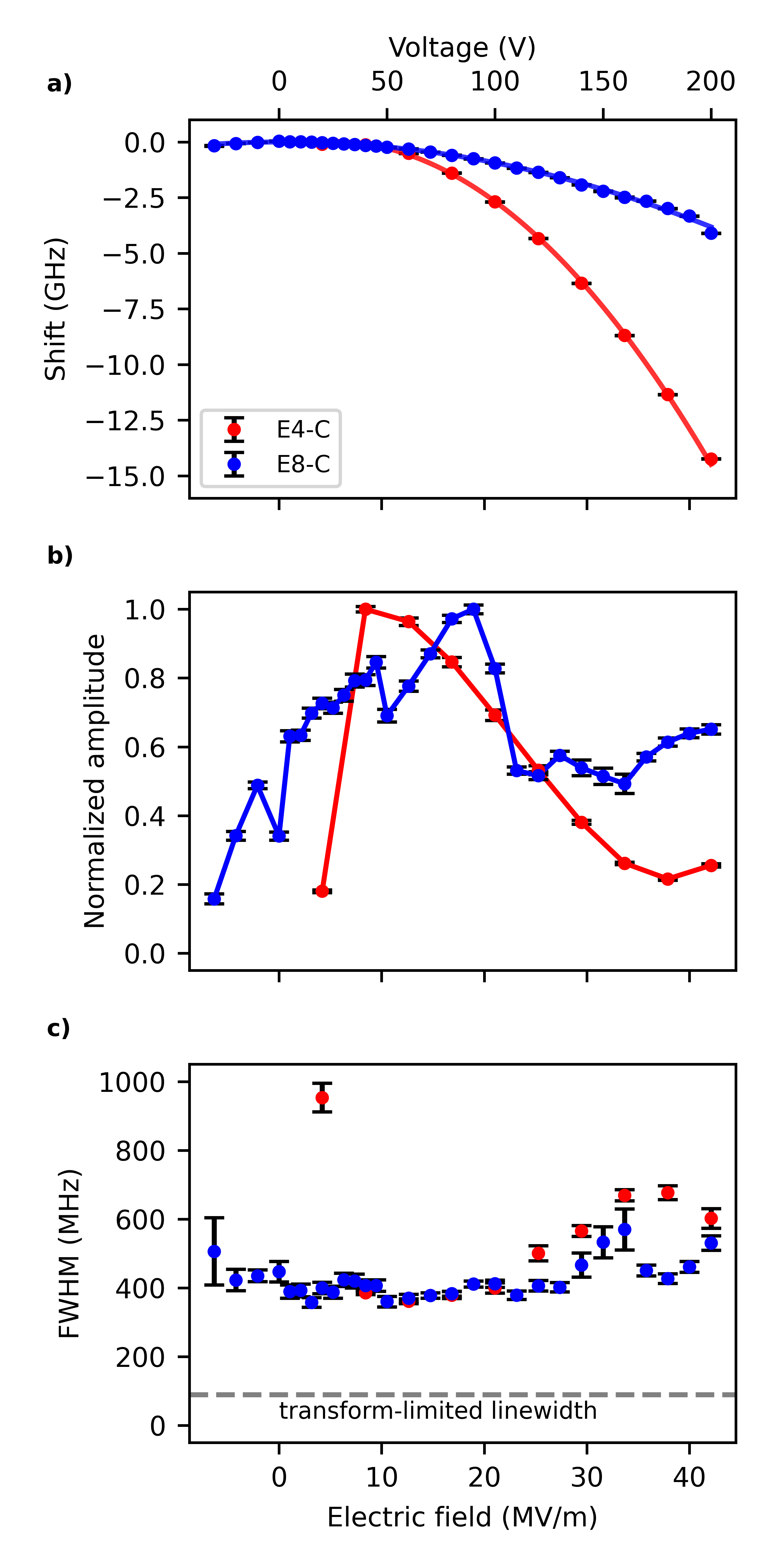}
\caption{\label{fig:Fig2}
\textbf{ Fitted figures of merit as a function of the applied electric field for 2 different single SiV$^-$ emitters. } 
Shown are the data for emitters labelled E4 (red) and E8 (blue), transition C (denoted E4-C and E8-C).  
a) Peak center frequency shifts $\Delta f$ with quadratic fit according to equations \ref{eq:tuning1} and \ref{eq:tuning2}.
b) Peak amplitudes. 
c) Full width at half maximum (FWHM) linewidths compared to the literature value of the transform-limited FWHM linewidth $1/(2\pi T_{\text{1,optical}}) \approx \SI{90}{\mega\hertz}$ \cite{Zuber2023_transformLimitedLinewidth,Zhou2017_lowLinewidthSiV}.
}
\end{figure}

Figure \ref{fig:Fig1} c) shows typical PL excitation spectra of an SiV$^-$ center as a function of the applied electric field.  We clearly observe a PL peak shift due to the DC Stark effect. Each spectrum is fitted using a Lorentzian to obtain
the center frequency $f(E)$, peak amplitude and peak linewidth as a function of the applied electric field. The resulting field-dependent data are presented in Figure \ref{fig:Fig2} a) confirming that the peak frequency shift $\Delta f$ is quadratic in the applied voltage or electric field $E$. As discussed above, this is expected since SiV$^-$ are inversion symmetric, leading to a vanishing linear contribution to the Stark shift. A second order shift is also observed in the Stark tuning of tin vacancies (SnV$^-$), which also exhibit inversion symmetry \cite{Aghaeimeibodi2021_Vuckovic_TinVac2021,DeSantis2021_Englund_StarkTun_TinVac_2021,bushmakin2025twophotoninterferencephotonsremote_Wrachtrup_tinVacancy}.
The data presented in Figure \ref{fig:Fig2} a) show two different SiV centers. Both show a quadratic Stark shift, but each clearly has a different polarizability.  One of the two emitters (E8-C) shifts by \SI{4}{\giga\hertz}, while the other (E4-C) shifts by \SI{14}{\giga\hertz} at the same applied local electric field.

In an earlier work, we showed that a variation in the applied voltage changes the SiV charge state \cite{rieger2024fast}. This process is induced by negatively biasing the electrode nearest to the SiV center, a procedure that promotes a transition to a dark state. Figure \ref{fig:Fig2} b) shows the PL intensity as a function of the applied field, exhibiting similar behavior as observed in \cite{rieger2024fast}: The peak practically vanishes when the nearest electrode is negatively biased and the amplitude is highest for positive voltages between $50$ and \SI{100}{\volt}.

We continue to discuss the linewidth as a function of the applied field. From the literature, we expect that the transform-limited full width at half maximum (FWHM) linewidth is $1/(2\pi T_{\text{1,optical}})=\SI{90}{\mega\hertz}$ \cite{Zuber2023_transformLimitedLinewidth,Zhou2017_lowLinewidthSiV}. Figure \ref{fig:Fig2} c) shows the fitted FWHM linewidths as a function of the applied electric field.
In the absence of an applied field, the linewidth is approximately \SI{400}{\mega\hertz}, increasing to values between $500$ and \SI{600}{\mega\hertz} towards high applied fields.
We expect that the additional line broadening is caused by a combination of the increasing induced static dipole moment that makes the center more sensitive to charge noise arising from current flow at higher applied voltages. Hence, the linewidth is expected to decrease by lowering the photocurrent caused by the off-resonant laser that stabilizes the SiV$^-$ charge state, e.g. by using a pulse sequence.

\begin{figure*}[!htb]
\centering
\includegraphics[scale=1]{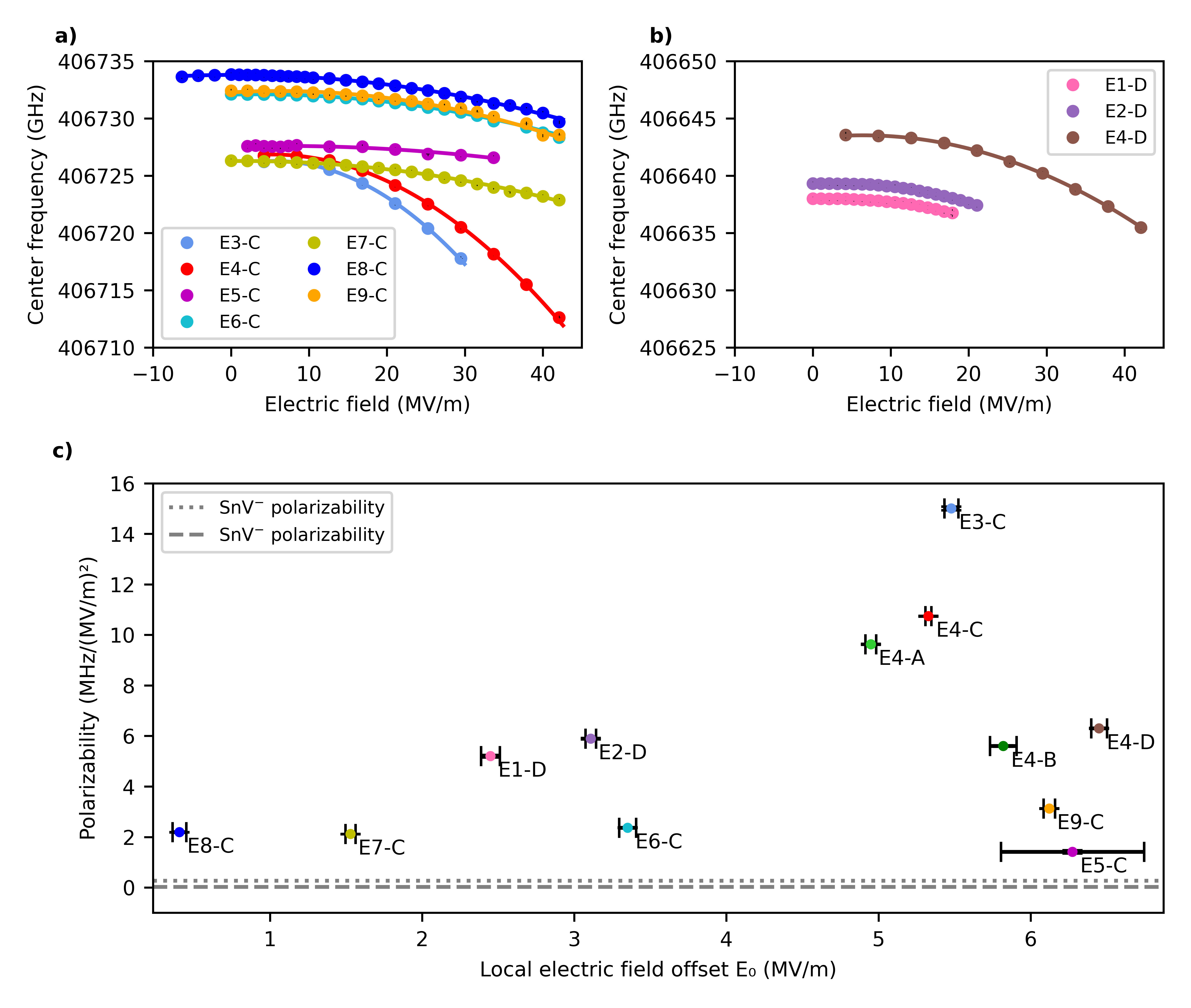}
\caption{\label{fig:Fig3}
\textbf{ Stark tuning of different individual emitters and statistics on analyzed parameters. }
The emitters are labelled from E1 to E9 and the measured transition is indicated with a letter from A to D as specified in Figure \ref{fig:Fig1} a).
a) Center frequencies of various emitters as a function of the applied electric field. Here, we show only the data that we measured for transition C.
b) Same as in a) but for transitions D.
c) Correlation plot of fitted polarizability versus offset electric field $E_0$. 
The data indicate that high polarizabilities are accompanied by high local electric field offsets. 
Tin vacancy polarizabilities are shown for comparison \cite{Aghaeimeibodi2021_Vuckovic_TinVac2021,DeSantis2021_Englund_StarkTun_TinVac_2021}.
}
\end{figure*}

We measured the PL voltage-dependent excitation spectra for 9 different emitters and observe center-to-center statistical variations of Stark tuning parameters. 
Figure \ref{fig:Fig3} a) summarizes all peak center frequencies measured for transition C (defined in Figure \ref{fig:Fig1}) of all centers and Figure \ref{fig:Fig3} b) the equivalent data for transition D. Figure \ref{fig:FigSI1} shows a plot of all measured center frequencies plotted as a function of the applied local electric field.
We clearly observe a trend that the optical transition frequencies at zero applied field vary, resembling the typical inhomogeneous distribution caused by locally varying strain and electric fields induced by crystal defects \cite{Evans2016_annealingProcedure}. The inhomogeneous distribution is $<\SI{10}{\giga\hertz}$ for both transitions C and D. This is expected for annealed bulk diamond and somewhat smaller than the inhomogeneous distribution of around \SI{15}{\giga\hertz} achieved for SiV$^-$ in optical nanostructures \cite{Evans2016_annealingProcedure}.

Figure \ref{fig:Fig3} c) shows the polarizability $\alpha$ as a function of the offset electric field $E_0$. 
The measured SiV$^-$ polarizabilites are significantly higher than those reported for tin vacancies \cite{Aghaeimeibodi2021_Vuckovic_TinVac2021,DeSantis2021_Englund_StarkTun_TinVac_2021,bushmakin2025twophotoninterferencephotonsremote_Wrachtrup_tinVacancy} and clearly vary between emitters. However, no direct correlation is observed between polarizability and zero-field frequency.
The offset electric field $E_0$ is always found to be positive and non-zero. This observation indicates that the spatial symmetry along the axis connecting the electrodes is broken. Furthermore, high polarizabilities occur preferentially alongside high local electric offset fields $E_0$.

\section{Discussion}

We observed center-to-center variations in polarizability and the local electric field offsets. We attribute the different offset electric field $E_0$ to the charged surroundings of individual SiV centers, including the space-charge region caused by band-bending at the nearby electrode and optical illumination \cite{goldblatt2024quantumelectrometrynonvolatilespace}, as well as the fluctuating charge environment caused by charged defects in the vicinity of each SiV$^-$. After identifying an emitter, we typically performed a single measurement that included also negative voltages. However, the photoluminescence excitation peak intensity was consistently found to be very weak below \SI{0}{\volt} or at least \SI{-20}{\volt}, consistent with our earlier work \cite{rieger2024fast}. The exact voltage dependence varies between emitters. Consequently, we measured primarily using positive voltages, potentially leading to asymmetries in the charging behavior. This may also explain the exclusively positive values of $E_0$, potentially since the optically induced space-charge region builds up slowly. Charged defects close to the SiV$^-$ can only explain the symmetry-breaking if they are a part of the optically induced space-charge region.

\sloppy Two aspects can explain the strong variation in the observed polarizabilities between \SI{1.4(1)}{} and \SI{15.0(1)}{\mega\hertz/(\mega\volt/\meter)^2}: Firstly, locally varying properties such as strain and Coulomb fields from charged defects near the SiV$^-$ can deform the wavefunction, hence altering the polarizability. It is known from literature that tin vacancies even show very strong linear contributions to the Stark shift, attributed to strain \cite{Aghaeimeibodi2021_Vuckovic_TinVac2021}. However, we do not observe significant linear Stark shifts for our SiV$^-$ emitters. Furthermore, strain may influence the response of NV$^-$ centers to electric fields \cite{lópezmorales2024quantumembeddingstudystrain}, a mechanism that could also translate to SiV.
Secondly, the variation of the polarizabilities may be an artifact arising from local electric fields that differ from the calculated values. This may arise due to screening of the applied electric field by an optically induced space charge region. Indeed, a study with a nitrogen-doped diamond found that space-charge regions built up during laser illumination and that they could screen the applied fields altogether at least up to a few \SI{}{\mega\volt/\meter} \cite{goldblatt2024quantumelectrometrynonvolatilespace}. However, we believe that the variation here is mainly due to strain and local electric fields.

Our results indicate that large polarizabilities tend to occur alongside high offset electric fields $E_0$.
Since high $E_0$ is caused by a strong space-charge screening of the applied electric field, our data contrasts with the hypothesis that high polarizabilities are an artifact caused by weaker screening. This leaves the hypothesis that locally varying properties such as strain and the Coulomb fields of other charged defects lead to the variation in polarizabilities. Further studies could measure all 4 optical transitions of each emitter, which would allow to calculate the strain using the known Hamiltonian \cite{Meesala2018_SiV_strainTuning} and to directly correlate the measured strain to the polarizability.

Each Stark shift that we measured can be fitted well by equation \ref{eq:tuning1}, in contrast to a recent report where certain tin vacancies exhibited a very noticable linear contribution to the Stark shift that was attributed to strong strain in the sample \cite{Aghaeimeibodi2021_Vuckovic_TinVac2021}.
In addition to the absence of a linear term, we note that higher order terms are not needed to fit the center frequencies, since the second order fits describe the data very well. Data on the quadratic fits of all measured individual Stark shifts can be found in Figure \ref{fig:FigSI2} in the Supplementary Information.

We attribute the substantially higher polarizability of SiV$^-$ compared to SnV$^-$ to differences in the confinement depths of the single-electron states. Density functional theory (DFT) calculations reveal that the $e_u$ states of SiV$^-$ are resonant with valence band maximum states and that they exhibit reduced localization \cite{GaliMaze}. This spatial delocalization enhances their size and polarizability and contributes to the larger Stark shifts observed in SiV$^-$ centers. In contrast, the $e_g$ states lie deep within the bandgap, resulting in strong localization and minimal contribution to the optical transition's Stark shift.

For heavier group-IV atoms, DFT calculations reveal that the $e_u$ states move deeper within the bandgap \cite{Thiering2018_MagnetoOpticalSpectra_G4V}, explaining the reduced polarizability observed in SnV$^-$ centers. This systematic trend across group-IV vacancy centers provides fundamental insight into the relationship between electronic structure and  characteristic electro-optical responses.
An additional comparison can be made to the recently studied inversion-symmetric nickel vacancy in diamond \cite{Morris2025}. The polarizability of this center is also significantly lower than that of the silicon vacancy and comparable to the polarizabilities of tin vacancies \cite{DeSantis2021_Englund_StarkTun_TinVac_2021,Aghaeimeibodi2021_Vuckovic_TinVac2021}. This is in agreement with our hypothesis above, as we do not expect one of the nickel vacancy energy levels to be resonant to the band edge states.

\section{Conclusion and Outlook}

Our results demonstrate the potential for electrical tuning of SiV$^-$ emission frequencies, for example to obtain indistinguishable photons for quantum repeaters. We find tuning ranges up to \SI{15}{\giga\hertz}, which is in the same order of magnitude as the width if the inhomogeneous optical emission frequency distribution caused by strain \cite{Evans2016_annealingProcedure}. 
We discuss the possibility that the SiV$^-$ polarizabilities are higher compared to those of tin or nickel vacancies because the $e_u$ level of the SiV$^-$ is resonant to valence band maximum states and thus more delocalized, making it weakly confined and more polarizable. Furthermore, our measurements show the presence of offset local electric fields and provide insights about their origin and the observed variations in the polarizability between different emitters. We attribute the offset electric fields to an optically induced space charge region and the variations in polarizability to local perturbations like strain and Coulomb fields of other defects.

Turning to possible directions for future investigations,  higher applied fields should significantly enhance the tuning range further and potentially allow frequency matching of arbitrarily selected SiV$^-$. Those higher fields could e.g. be realized by electrodes with smaller separations of $1$ to \SI{4}{\micro\meter}.
Significantly stronger electric fields could lead to charge state instability, necessitating mitigation strategies for charge-state stabilization and charge-induced decoherence. Rapid voltage modulation protocols, as demonstrated in our previous work \cite{rieger2024fast}, could enable charge state initialization before Stark tuning, potentially extending the accessible tuning range. Complete characterization of the Stark tensor through independent control of electric fields in all spatial dimensions would provide a comprehensive understanding of the electro-optical coupling. Finally, investigating the impact of rapidly varying electric fields on excited state dynamics may reveal new opportunities for photon shaping and quantum state manipulation.

\section*{Materials and methods}

\subsection*{Sample preparation}\label{sec:sample}

Silicon vacancy centers were created by implanting silicon into electronic grade, single-crystal diamond produced by Element Six. Element Six grew the diamond using chemical vapor deposition (CVD) and diced it into dimensions of \SI{3}{\milli\meter} $\times$ \SI{3}{\milli\meter} $\times$ \SI{0.5}{\milli\meter}. This grade of diamond typically contains nitrogen and boron impurity levels below \SI{5}{ppb} and \SI{0.5}{ppb}, respectively \cite{E6_handbook}. 

CuttingEdge Ions implanted $^{28}$Si ions at \SI{132}{\kilo\electronvolt} with an expected implantation depth of \SIrange{100}{150}{\nano\meter} and a dose of \SI{5e9}{ions/\centi\meter^{-2}}. After implanatation, the samples were annealed in multiple steps: \SI{4}{\hour} ramp to \SI{400}{\celsius}, \SI{8}{\hour} at \SI{400}{\celsius}, \SI{12}{\hour} ramp to \SI{800}{\celsius}, \SI{8}{\hour} at \SI{800}{\celsius}, \SI{12}{\hour} ramp to \SI{1100}{\celsius}, and \SI{10}{\hour} at \SI{1100}{\celsius}, all under high vacuum ($<\SI{1e-7}{mbar}$) \cite{Evans2016_annealingProcedure}. During this process,  the temperature is gradually incremented to maintain high vacuum within the annealing chamber, essential for preventing surface graphitization of the diamond \cite{ruf2019optically}.

The sample was cleaned sequentially in aqua regia (\SI{1}{\hour} at \SI{90}{\celsius}) and concentrated sulfuric acid (\SI{1}{\hour} at \SI{225}{\celsius}), followed by oxygen plasma exposure to achieve oxygen termination. We fabricated interdigitated electrodes with \SI{7.6}{\micro\meter} spacing using optical lithography and metal evaporation starting with \SI{20}{\nano\meter} of titanium followed by \SI{200}{\nano\meter} of gold. Each electrode connects to \SI{200}{\micro\meter} $\times$ \SI{200}{\micro\meter} bond pads. Finally, we mounted the sample on a well-thermalized copper plate and wire-bonded it to a custom-printed circuit board chip carrier.

\subsection*{Optical and electrical measurements}

The sample was cooled to \SI{5}{\kelvin} using an attoDRY800 closed-cycle cryostat from attocube systems AG.
Electrical voltage was applied using a Keithley 2450 source-measure unit, which also measured currents. Both the Keithley and half of the electrodes were grounded to the cold base plate of the cryostat in a star-shaped configuration.

Optical measurements were performed using a custom-built confocal microscope setup. Off-resonant excitation employed Toptica Photonics SE iBeam smart diode lasers with either \SI{740}{\tera\hertz} (\SI{405}{\nano\meter}) or \SI{582}{\tera\hertz} (\SI{515}{\nano\meter}) optical frequency or a Laser quantum LLC torus laser with an optical frequency of \SI{564}{\tera\hertz} (\SI{532}{\nano\meter}).

For resonant excitation, we used a narrow-linewidth continuous-wave laser (C-Wave GTR from Hübner Photonics), wavelength-stabilized to a wavemeter with \SI{60}{\mega\hertz} absolute accuracy and \SI{2}{\mega\hertz} resolution (High Finesse WS7-60 IR-I). Power stabilization was achieved using a software PID controller and analog modulation via a fast acousto-optic modulator (\SI{15}{\nano\second} rise time, driven at \SI{350}{\mega\hertz} in the first diffraction order). 

All excitation lasers were coupled into single-mode fibers for spatial mode filtering.
After combining both excitation lasers with a free-space dichroic mirror, a \SI{400}{\tera\hertz} (\SI{750}{\nano\meter}) frequency-long pass filter removed optical noise and fiber-induced Raman signal frequencies from the excitation laser spectra.
Laser focusing was accomplished using an $f=\SI{2.87}{\milli\meter}$, $0.82$ NA, low-temperature apochromatic objective (LT-APO/VISIR/0.82 from attocube systems AG).

Detection employed single-mode fiber-coupled avalanche photodiodes (APDs) with approximately \SI{700}{\hertz} dark counts or superconducting single-photon detectors from Single Quantum with $<\SI{1}{\hertz}$ dark counts. 
Three \SI{400}{\tera\hertz} (\SI{750}{\nano\meter}) frequency-short pass filters and one \SI{545}{\tera\hertz} (\SI{550}{\nano\meter}) frequency-short pass filter separated the excitation laser from the SiV$^-$ phonon sideband signal. A \SI{375}{\tera\hertz} (\SI{800}{\nano\meter}) frequency-long pass filter removed the diamond Raman signal from the \SI{407}{\tera\hertz} (\SI{737}{\nano\meter}) resonant excitation laser, which appears at \SI{367}{\tera\hertz} (\SI{816}{\nano\meter}). 

\begin{acknowledgments}

This work was supported by BMFTR through project epiNV (13N15702), SPINNING (13N16214) and by the Bayerisches Staatsministerium für Wissenschaft und Kunst through project IQSense via the Munich Quantum Valley (MQV).

We gratefully acknowledge support from the Deutsche Forschungsgemeinschaft (DFG, German Research Foundation) under projects PQET (INST 95/1654-1) and Germany’s Excellence Strategy – EXC-2111 – 390814868.

We used the artificial intelligence Claude from Anthropic PBC to review the understandability of our manuscript.  Moreover, we acknowledge the use of ChatGPT and Claude for generating python code snippets used for experimental control software, as well as the analysis and processing of experimental data for publication.

\end{acknowledgments}

\bibliography{main}

\newpage

\appendix
\renewcommand{\thefigure}{S\arabic{figure}}
\setcounter{figure}{0}  %

\section{Supplementary Information}

\begin{figure*}[h]
\centering
\includegraphics[scale=1]{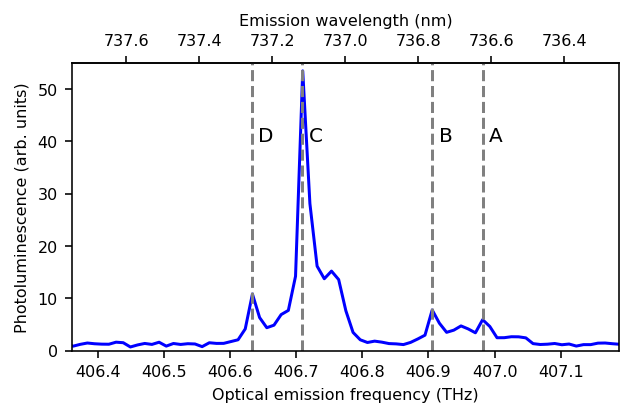}
\caption{\label{fig:FigPL}
\textbf{Photoluminescence spectrum of SiV$^-$,} 
which used a \SI{564}{\tera\hertz} (\SI{532}{\nano\meter}) laser for excitation and an $f=\SI{500}{\milli\meter}$ spectrometer with a grating that has $1200$ lines per millimeter for detection. We assign lines A, B, C and D to one of the emitting SiV$^-$. The fact that there are more emission lines indicates that multiple SiV$^-$ with varying degrees of strain contribute. 
}
\end{figure*}

Figure \ref{fig:FigPL} shows a representative photoluminescence spectrum that we used to make an initial estimation of the optical transition frequencies of the SiV$^-$. We see the typical SiV$^-$ spectrum with lines A, B, C and D as described in Figure \ref{fig:Fig1} a). For the particular SiV$^-$ monitored here, the ground state splitting is \SI{76}{\giga\hertz} and the excited state splitting is \SI{273}{\giga\hertz}. 
The spectrum shows more than these 4 lines and thereby indicates the presence of more than just one single SiV$^-$ center. The centers are also strained to varying degrees. We did not attribute the other transitions for the sake of clarity of the figure. The SiV$^-$ density in the sample is too high to spatially resolve single SiV$^-$ in confocal micro-photoluminescence measurements. However, it is low enough to clearly distinguish different emitters spectrally using confocal micro-photoluminescence excitation spectroscopy with high spectral resolution which we used for the measurements in the main text.

\begin{figure*}[h]
\centering
\includegraphics[scale=1]{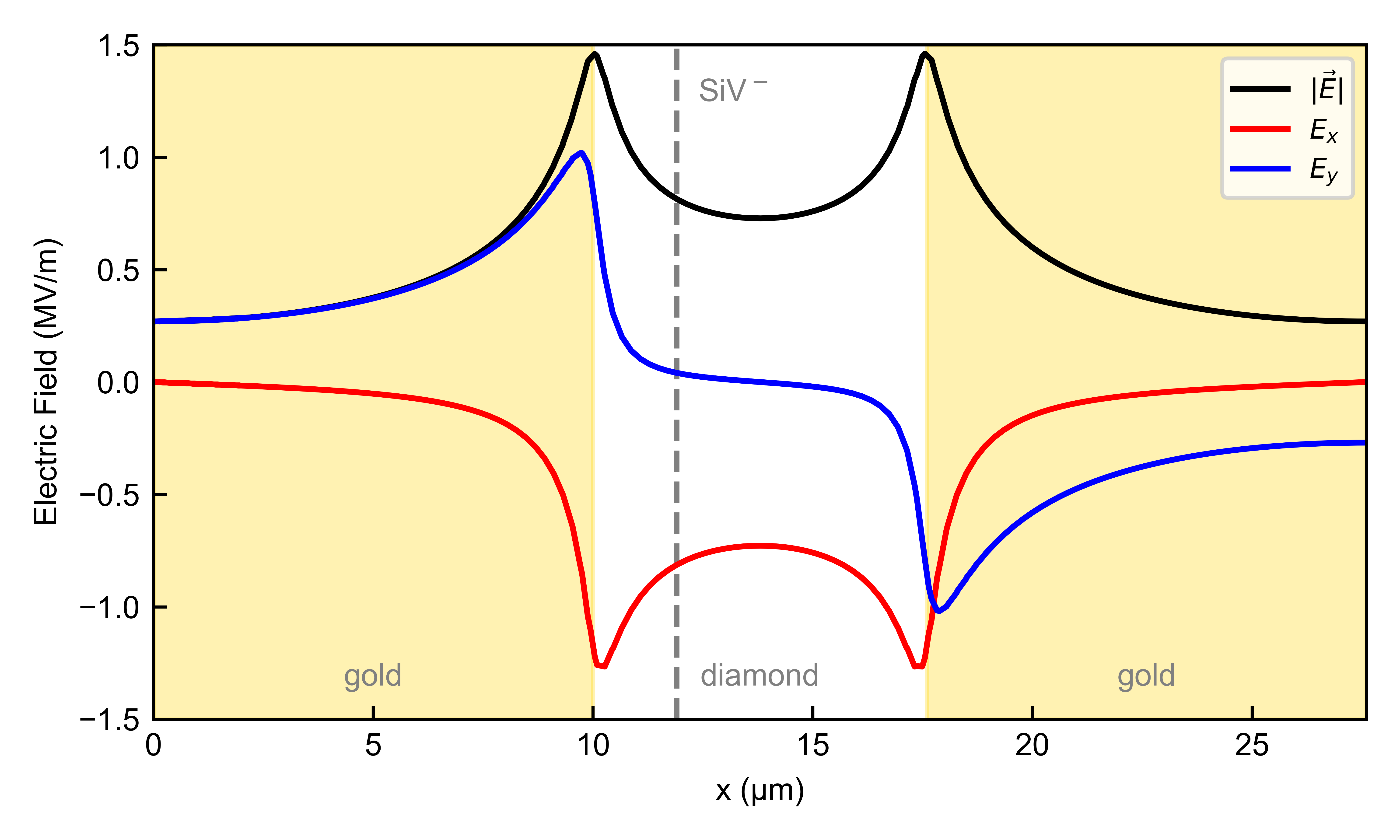}
\caption{\label{fig:FigSICOMSOL}
\textbf{COMSOL simulation of the static electric field.} 
The field is calculated in the diamond \SI{100}{\nano\meter} (resembling SiV$^-$ implantation depth) below the surface along a line that runs perpendicular to the edges of the electrode fingers. In the simulation, we applied \SI{10}{\volt} to the right gold electrode and grounded the left one. The electrode separation is \SI{7.6}{\micro\meter}. The x-direction runs parallel to the diamond surface between the electrodes. Positive y-direction is from the diamond surface upwards into the vacuum. The dashed line indicates the approximate position of the SiV$^-$ that we studied.
}
\end{figure*}

Figure \ref{fig:FigSICOMSOL} shows the results of the COMSOL simulation of the electric field components $E_x$, $E_y$ and the norm $|\Vec{E}|$ as a function of the x-coordinate between two electrode fingers. The applied voltage is \SI{10}{\volt} and the field was calculated \SI{100}{\nano\meter} below the surface, which is the expected implantation depth. The electrode separation is \SI{7.6}{\micro\meter} and the electrodes are \SI{10}{\micro\meter} wide and \SI{150}{\nano\meter} thick. Pure vacuum surrounds the diamond and the electrodes. 

\begin{figure}[h]
\centering
\includegraphics[scale=1]{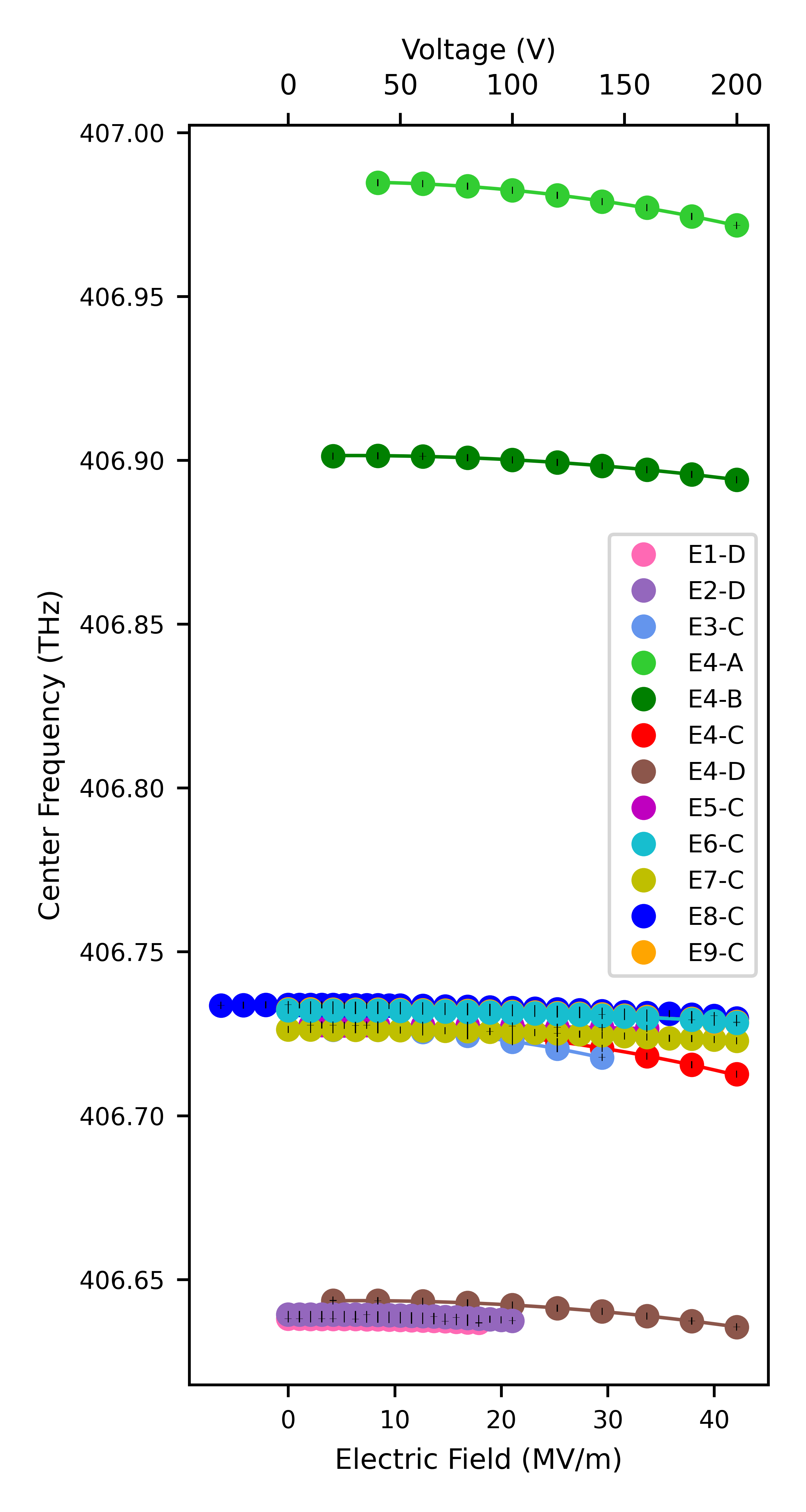}
\caption{\label{fig:FigSI1}
\textbf{All measured peak center frequencies as a function of applied voltage and applied local electric field.}
The center frequencies are clustered into 4 regions which correspond to the 4 optical transitions allowed in SiV$^-$ at zero magnetic field. 
The voltage range over which peak fitting could be performed to determine center wavelengths varied between individual emitters, reflecting voltage-dependent intensity variations arising from charge state conversion processes.
}
\end{figure}

\begin{figure*}[h]
\centering
\includegraphics[scale=1]{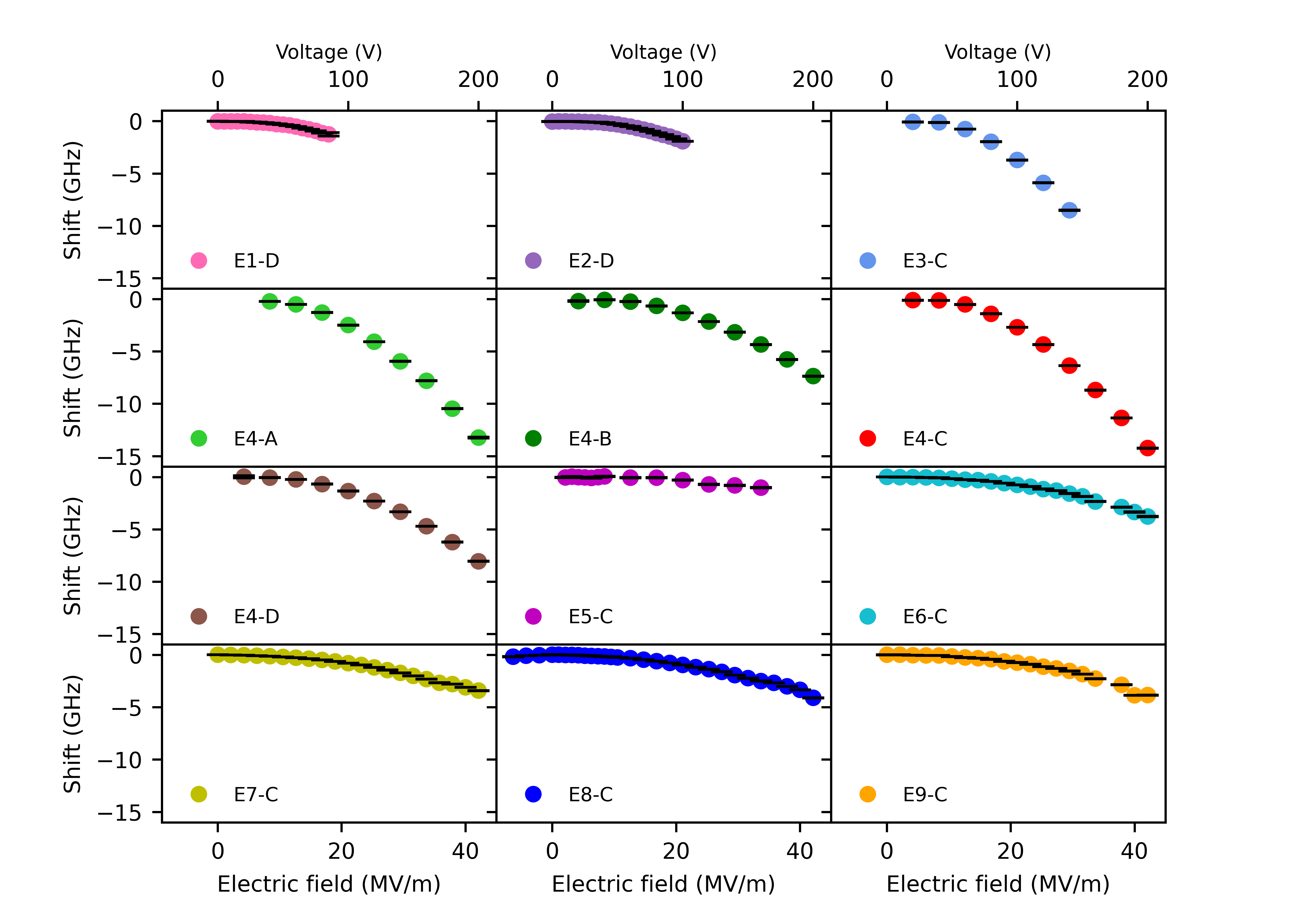}
\caption{\label{fig:FigSI2}
\textbf{ Overview of all measured peak Stark shifts as a function of applied voltage and applied local electric field.}
The polarizability of the second-order Stark tuning clearly varies between emitters, with some of the emitters reaching Stark shifts of $\approx\SI{10}{\giga\hertz}$ while some reach only a few \SI{}{\giga\hertz}.
}
\end{figure*}

Figure \ref{fig:FigSI1} displays the absolute peak center frequencies of all measured SiV$^-$ as a function of the applied voltage and applied local electric field. For emitter E4, we measured all 4 optical transitions. 
In Figure \ref{fig:FigSI2} we present the Stark shifts $\Delta f$ for the same data, which are the deviation from the maximum measured frequency $f_{\text{max}}$. Note that the peaks shift to lower transition frequencies. 
In addition to the change of transition frequency with applied electric field, the peak amplitudes also change. This is caused by charge state conversion \cite{rieger2024fast}. At low and at elevated fields, this can lead to peak intensities in the noise floor, so that we could not fit the peaks at these applied voltages.

\end{document}